# Magnetic Moment Formulas of Baryons Determined by Quantum Numbers

*Yi-Fang Chang*
*Department of Physics, Yunnan University, Kunming 650091, China*
(E-mail: yifangchang1030@hotmail.com)

**Abstract**: We propose that the magnetic moment formulas of baryons may be determined by quantum numbers, and obtain three formulas: $\mu = \{\mu_0 + aQ(Q+1) + bU(U+1)I + cS\}\mu_N$, $\mu = \{\mu_0 + aQ(Q+1) + bU(U+1) + cS\}\mu_N$ and $|\mu| = \{A[Q(Q+1) + \frac{S}{2}] + B(\overline{U}+1)\}\mu_N$, in which $\overline{U} = I + \frac{1}{2}(|Y| - |Q|)$ is a composed quantum number. This is a new type of magnetic moment formula, and agrees better with the experimental values. It is also similar to corresponding mass formulas of hadrons.

Usually calculated methods of magnetic moment for baryons are based on various quark models. From this a well-known result $\mu_p/\mu_n = -3/2$ [1] and some progresses have been obtained. But, the data of magnetic moments do not agree with original quark model [2] except $\mu(\Xi^0) = 2\mu(\Lambda)$, and these methods connects necessarily with the structure and internal interactions of particles. Therefore, the magnetic moment formula of baryons is still an opinions vary, and a researched question from various aspects [3-31], which included quark mass and rotation, QCD, lattice, field theory, and string, etc. Some models are very complex.

Applied an analogous method of mass formulas of hadrons determined by quantum numbers, for example, the Gell-Mann-Okubo mass formula, we propose a new type of magnetic moment formula for baryons determined by quantum numbers.

First, we propose a magnetic moment formula:

$$\mu = \{\mu_0 + aQ(Q+1) + bU(U+1)I + cS\}\mu_N, \quad (1)$$

in which $\mu_N = e\hbar/2m_p$ is a magnetic moment of proton. It agrees qualitatively with data except $\Sigma^0$, when $\mu_0 = -0.943$, *a*=2.05, *b*=-0.9, *c*=0.33.

Next, the second magnetic moment formula is:

$$\mu = \{\mu_0 + aQ(Q+1) + bU(U+1) + cS\}\mu_N. \quad (2)$$

It agrees better with data, when $\mu_0 = -0.943$, *a*=2.05, *b*=-0.485, *c*=0.33.

Third, we introduce a composed quantum number

$$\overline{U} = I + \frac{1}{2}(|Y| - |Q|), \qquad (3)$$

which is namely the values of U-spin for U($\Lambda$)=0 and U($\Sigma^0$)=1. Such the magnetic moment of the $J^p = 1^+/2$ baryon octet may classify four groups: a $\overline{U}$ =0 singlet ($\Lambda$), two $\overline{U}$ =1/2 doublet (p, $\Sigma^+$) and ($\Xi^-$, $\Sigma^-$), a $\overline{U}$ =1 triplet ($\Xi^0$, $\Sigma^0$, n). The experimental values seem to show that various magnetic moments of baryon octet are classified according to two quantum numbers charge Q and $\overline{U}$. Then assume that the magnetic moments are positive for baryons with Q=1, while the magnetic moments are negative for baryons with Q=0 or –1. So we propose an absolute value formula of magnetic moment for baryon [32]:

$$|\mu| = \{A[Q(Q+1) + \frac{S}{2}] + B(\overline{U}+1)\}\mu_N. \qquad (4)$$

The magnetic moments take also absolute value in quark model [2]. Let A=0.679 and B=0.9565, we obtain the theoretical values. In the table 1 we compare the experimental values [33] with some theoretical results.

Table 1. Comparison of the experimental values of magnetic moments with some theoretical results

| $\mu(B)$ | Ref.[10] | Ref.[11] | Ref.[12] | Ref.[14] | Ref.[20](QCDSA) | Ref.[21] | Ref.[27](NQM)] |
|---|---|---|---|---|---|---|---|
| p | 2.65 | 2.79 | input | 2.79 | 2.54 | 2.34 | 2.79 |
| n | -1.90 | -1.92 | -1.86 | -1.91 | -1.69 | -1.61 | -1.86 |
| $\Lambda$ | -0.60 | -0.61 | input | -0.61 | -0.69 | -0.58 | -0.61 |
| $\Sigma^+$ | 2.36 | 2.46 | 2.68 | 2.48 | 2.48 | 2.25 | 2.68 |
| $\Sigma^0$ |  |  | -1.61 | -1.53 | -0.8 | -0.67 | -1.61 |
| $\Sigma^-$ | -1.20 | -1.20 | -1.04 | -0.98 | -0.90 | -0.89 | -1.04 |
| $\Xi^0$ | -1.23 | -1.20 | -1.43 | -1.25 | -1.49 | -1.28 | -1.44 |
| $\Xi^-$ | -0.72 | -0.73 | -0.50 | -0.50 | -0.63 | -0.50 | -0.51 |

| $\mu(B)$ | Ref.[36] O+D | Formula(1) | Formula(2) | Formula(4) | Expt. [33] |
|---|---|---|---|---|---|
| p | 2.793 | 2.793 | 2.793 | 2.7928 | 2.7928 |
| n | -1.913 | -1.913 | -1.913 | -1.9130 | -1.9130 |
| $\Lambda$ | -0.613 | -0.613 | -0.613 | -0.6170 | -0.613 $\pm$ 0.004 |
| $\Sigma^+$ | 2.458 | 2.759 | 2.463 | 2.4533 | 2.458 $\pm$ 0.010 |
| $\Sigma^0$ |  | -2.553 | -1.583 | -1.5735 | -1.61 $\pm$ 0.08 |
| $\Sigma^-$ | -1.160 | -1.341 | -0.977 | -1.0953 | -1.160 $\pm$ 0.025 |
| $\Xi^0$ | -1.25 | -1.253 | -1.253 | -1.2340 | -1.250 $\pm$ 0.014 |
| $\Xi^-$ | -0.651 | -0.647 | -0.647 | -0.7558 | -0.6507 $\pm$ 0.0025 |

Moreover, there is a simple approximate relation between two constants in (4): B/A=1.409

$\approx -(\mu_p / \mu_n) = 1.460 \approx 3/2$. Further, many relations of baryon magnetic moments may be obtained based on the relations of quantum numbers. For instance, from the formula (4) we may derive following results:

$$2[\mu(\Lambda) + \mu(\Sigma^0)] = \mu(n) + 2\mu(\Xi^0) \quad (-4.446 \approx -4.413),$$

$$3\mu(\Lambda) + \mu(\Sigma^0) = \mu(n) + 2\mu(\Xi^-) \quad (-3.449 \approx -3.214),$$

$$\mu(n) + \mu(\Xi^0) = 2\mu(\Sigma^0) \quad (-3.163 \approx -3.22),$$

$$\mu(\Lambda) + \mu(\Xi^0) = \mu(\Sigma^-) + \mu(\Xi^-) = 3\mu(\Lambda) \quad (-1.863 \approx -1.811 \approx -1.839),$$

$$\mu(p) + \mu(\Sigma^0) = \mu(\Sigma^+) + \mu(\Xi^0) \quad (-1.1828 \approx -1.208),$$

$$|\mu(p)| + |\mu(\Xi^0)| = |\mu(\Sigma^+)| + |\mu(\Sigma^0)| \quad (4.0428 \approx 4.068), \text{ etc.}$$

Perhaps, the magnetic moments of antiparticles are only opposite values with corresponding particles, for example, $\mu(\overline{p}) = -2.8005(90)\mu_N = -\mu(p)$ for negative proton [34].

Linde, Puglia and Dahiya, et al., discussed magnetic moments of 3/2 resonances and decuplet [35-37]. When the magnetic moment formula (4) is extended to the $J^P = 3^+/2$ baryon decuplet, it should add a spin-term, is namely:

$$|\mu| = \{A[Q(Q+1) + \frac{S}{2}] + B(\overline{U}+1) + CJ(J-\frac{1}{2})\}\mu_N. \quad (5)$$

This is also similar to the mass formula of corresponding baryons. If A and B are the same values, we will obtain C=1.0692 since $\mu(\Omega^-)$=-2.02±0.05 [34]. Then other magnetic moments of decuplet baryons can be predicted:

Table 2. A prediction on magnetic moments of decuplet baryons

|  | $\Delta^{++}$ | $\Delta^+$ | $\Delta^0$ | $\Delta^-$ | $\Sigma^{*+}$ |
|---|---|---|---|---|---|
| $\overline{U}$ | 1 | 3/2 | 2 | 3/2 | 1/2 |
|  | 6A+2B+1.5C | 2A+2.5B+1.5C | 3B+1.5C | 2.5B+1.5C | 1.5A+2.5B+1.5C |
|  | 9.2558 | 5.3531 | -4.4733 | -3.9951 | 3.7516 |

| $\Sigma^{*0}$ | $\Sigma^{*-}$ | $\Xi^{*0}$ | $\Xi^{*-}$ |
|---|---|---|---|
| 1 | 1/2 | 1 | 1/2 |
| -0.5A+2B+1.5C | -0.5A+1.5B+1.5C | -A+2B+1.5C | -A+1.5B+1.5C |
| -3.1773 | -2.6989 | -2.837 | -2.3596 |

Of course, in formula (5) A and (or) B are also probably different. Or combined spin J, the quantum number $\overline{U}$ may be extended.

In a word, it is a new type of magnetic moment formula, which is independent to the constituents, structures and interactions inside baryons, and is similar to corresponding mass formulas of hadrons.